%% file: main.tex
\documentclass[prd,showpacs,showkeys,preprintnumbers,aps,twocolumn]{revtex4}
\usepackage{graphicx}% Include figure files
\usepackage{dcolumn}% Align table columns on decimal point

\input{def}
  \def\S(#1,#2){{S^{#1}_{#2}}}
 \def\Su(#1,#2){{\hat{S}^{#1}_{#2}}}
 \def\Sd(#1,#2){{\check{S}^{#1}_{#2}}}
 \def\C(#1){C_{#1}}
  \def\CG(#1){(C\gamma_5)_{#1}}
  \def\G(#1){{\Gamma}^{#1}}
  \def\g{{\gamma}}  
  \def\q{{\bar q}}

  \def\s{{\bar s}}

   \def\Ci{C^{-1}}

  \def\gt_#1{(-C\g_{#1}\Ci)}
  \def\bra{\langle}
  \def\ket{\rangle}
  \def\qcond{\bra 0|\q q|0\ket}

  \def\sc{\bra 0|\s s|0\ket}

  \def\psla{p{\raise1pt\hbox{$\!\!/$}}}
  \def\qsla{q{\raise1pt\hbox{$\!\!\!/$}}}
  \def\knd(#1,#2){\delta_{#1#2}}

  \def\sGs{g\langle 0|\bar s\sigma^{\mu\nu}(\lambda^a/2)G^a_{\mu\nu}s|0\rangle}

  \def\GG{\langle 0|{\alpha_s\over\pi}G^{a\mu\nu}G^{a}_{\mu\nu}|0\rangle}
  
  \def\vecp{\textnormal{\mathversion{bold}$p$}}

\begin{document}

\preprint{TH-991}

\title{$\Theta^{++}$ from QCD sum rule}
%\title{$J^P=3/2^-$, $I=1$ pentaquark from QCD sum rule}

\author{Tetsuo NISHIKAWA}
\email{nishi@post.kek.jp}
\author{Yoshiko KANADA-EN'YO}
\email{yoshiko.enyo@kek.jp} 
\affiliation{%
Institute of Particle and Nuclear Studies, 
High Energy Accelerator Research Organization, 1-1, Ooho, 
Tsukuba, Ibaraki, 305-0801, Japan
}
\author{Yoshihiko KONDO}
\email{kondo@kokugakuin.ac.jp}
\affiliation{%
Kokugakuin University, Higashi, Shibuya, Tokyo 150-8440, Japan
} 
\author{Osamu MORIMATSU}
\email{osamu.morimatsu@kek.jp}
\affiliation{%
Institute of Particle and Nuclear Studies, 
High Energy Accelerator Research Organization, 1-1, Ooho, 
Tsukuba, Ibaraki, 305-0801, Japan
}

%\date{April 24, 2003}
\date{\today}
\begin{abstract}
We study the pentaquark $uudd\bar s$ with $J=3/2$ and $I=1$ ($\Theta^{++}$) in the QCD sum rule approach.
%If such a state exists and lies below the $\Delta-K$ threshold, it can be extremely narrow  
%since its decay into $K-N$ state is strongly suppressed due to the $D$-wave centrifugal barrier.
We derive the QCD sum rules for positive and negative parity states
of the pentaquark. 
The QCD sum rule predicts that 
there exists $\Theta^{++}$ with negative parity
and its mass is $1.5\sim1.6$ GeV.
The negative parity $\Theta^{++}$ can be extremely narrow, since it lies much below the $\Delta K$ threshold and the decay into $KN$ state is strongly suppressed due to the $D$-wave centrifugal barrier.
Also, the possibility of the existence of the $\Theta^{++}$ with positive parity is not excluded. Although it nearly degenerates with the negative parity state, 
it may be broader than the negative parity state.  
\end{abstract}
\maketitle

%%%%%%%%%%%%%%%%%%%%%%%%%%%%%%%%%%
%\newpage
An exotic baryon state with positive strangeness, $\Theta^{+}$, has been recently observed by LEPS collaboration in Spring-8 \cite{leps} and the subsequent experiments \cite{diana,clasa,clasb,saphir,itep,hermes,itep-2,zeus,clas-c}.
The mass of the $\Theta^{+}$ is $\sim1540\,{\rm MeV}$ and the width is unusually small: $\Gamma<25\,{\rm MeV}$.
This state cannot be a three-quark state since it has positive strangeness, 
and therefore the minimal quark content is $(uudd{\bar s})$.

The spin and the parity have not yet been experimentally determined, 
while there is an experimental indication that
$\Theta^{+}$ has $I=0$ \cite{clasa,saphir}.
In order to clarify the quantum numbers and to
understand the structure of the $\Theta^{+}$,
intense theoretical studies have been done so far \cite{jaffe,capstick,karliner,sasaki,csikor,zhu,matheus,sugiyama}.
On the other hand, it has been suggested that there exist various pentaquark states with unnatural spin, isospin and parity \cite{ellis,borisyuk,amdpenta},
which indicates that the level structure of pentaquark may be quite different from those expected from the ordinary hadron spetra.

Among various states newly predicted in Ref.\cite{ellis,borisyuk,amdpenta}, we focus here on the $J^P=3/2^-$ and $I=1$ pentaquark,
which belongs to a new family of flavor SU(3), the 27-pret.
This state has been predicted by a quark model to exist as a low lying state for $uudd\bar s$ system and 
nearly degenerate with $J^P=1/2^+$ or $3/2^+$ and $I=0$ pentaquark \cite{amdpenta}.
If such a state lies much below the $\Delta K$ threshold, it can be extremely narrow, since it decays only to $D$-wave $KN$ states and the width is strongly suppressed due to the high centrifugal barrier \cite{amdpenta}.
The $J^P=3/2^-$ and $I=1$ state can be a candidate of new pentaquarks that are enough narrow to be observed.
To study this state theoretically would help experimental search for the new particle, $\Theta^{++}$.

In order to ascertain 
%whether the pentaquark is extremely narrow,
the existence of the narrow pentaquark state with $J^P=3/2^-$ and $I=1$,
it is crucial to estimate its absolute mass,
since the width is sensitive to the energy difference from the $\Delta K$ threshold.
In this paper, 
we study $J=3/2$ and $I=1$ pentaquark ($\Theta^{++}$) by using the method of QCD sum rule \cite{SVZ}, which is closely related to the fundamental theory and able to evaluate the absolute masses of hadrons without any model assumptions.
In QCD sum rule approach, a correlation function of an interpolating field is calculated by the use of 
the operator product expansion (OPE),
and is compared with the spectral representation via dispersion relation.
The sum rules relate hadron properties to the vacuum expectation values
of QCD operators (condensates), such as $\langle 0|{\bar q}q|0\rangle$, $\langle 0|(\alpha_{s}/\pi)G^2|0\rangle$ and so on.

The correlation function from which we derive the QCD sum rule is
\bey
\Pi_{\mu\nu}(p)=-i\int d^4x \exp(ipx)
\langle 0|T\left[\eta_{\mu}(x)\bar\eta_{\nu}(0)\right]|0\rangle,
\label{corfun}
\eey
where ${\eta}_{\mu}$ is an interpolating field for the pentaquark with $J^P=3/2$ and $I=1$.
%There are various ways of the construction of ${\eta}_{\mu}$.
We use the following interpolating field,
\bey
{\eta}_{\mu}=\epsilon_{cfg}
(\epsilon_{abc}u_{a}^{T}C\gamma_5 d_{b})
(\epsilon_{def}u_{d}^{T}C\gamma_{\mu}u_{e})
C\bar{s}_{g}^{T},
\label{current}
\eey
where $u$, $d$ and $s$ are up, down and strange quark fields, resepectively, roman indices $a,b,\ldots$ are color, $C$ denotes charge conjugation matrix, and $T$ transpose. 
$\epsilon_{abc}u_{a}^{T}C\gamma_5 d_{b}$ is a color ${\bar 3}$ scalar diquark operator with $I=0$.
$\epsilon_{def}u_{d}^{T}C\gamma_{\mu}u_{e}$ is a color ${\bar 3}$ axial-vector diquark operator with $I=1$.
Thus \eq{current} is totally $I=1$ and it contains the state with $J^P=3/2^-$.
The way of constructing the interpolating field, \eq{current}, is based on 
the picture of the pentaquark structure found from the quark model calculation mentioned above \cite{amdpenta}.
According to the Ref.\cite{amdpenta}, the pentaquark with $J^P=3/2^-$ and $I=1$ consists of two color ${\bar 3}$ diquarks and an anti-strange quark.
One of the diquarks has $S=0$ and $I=0$ and the other $S=1$ and $I=1$.
Evidently, the interpolating field, \eq{current}, possesses the same diquark structure.

The correlation function, \eq{corfun}, has various tensor structures,
\bey
\Pi_{\mu\nu}(p)&=&g_{\mu\nu}\psla\Pi_{1}(p^2)+g_{\mu\nu}\Pi_{2}(p^2)\cr
&&+\g_\mu \g_\nu \Pi_{3}(p^2)+\cdots.
\eey
We are interested in the terms proportional to $g_{\mu\nu}$:
\bey
\Pi(p)\equiv\psla\Pi_{1}(p^2)+\Pi_{2}(p^2),
\label{3/2part}
\eey
since these terms receive the contribution of pure $J=3/2$ states.
In the other terms, $J=1/2$ states contribute as well as $J=3/2$ states \cite{ioffe}.

We can relate the correlation function with the spectral function via Lehman representation, 
\begin{eqnarray}\label{Lehman}
\Pi(p_0,\vecp
)=\int_{-\infty}^{\infty}{\rho(p'_0,\vecp)\over p_0-p'_0}dp'_0,
\end{eqnarray}
where $\rho(p_0,\vecp)$ is the spectral function.
% given by
%$\rho(p_0,\vecp)=-(1/\pi){\rm Im}\Pi(p_0,\vecp)$.
On the other hand, in the deep Euclid region, $p_0^2\rightarrow-\infty$, 
the correlation function can be evaluated by an operator product expansion. Then the correlation function is expressed as a sum of various vacuum condensates.
Using the analyticity, we obtain a relation between the imaginary part of the correlation function evaluated by an OPE, $\rho^{\rm OPE}$, and the spectral function as
\begin{eqnarray}
\int_{-\infty}^{\infty}dp_0\rho^{\rm OPE}(p_0,\vecp)W(p_0)
=\int_{-\infty}^{\infty}dp_0\rho(p_0,\vecp)W(p_0),
\label{QSR}
\end{eqnarray}
where $W(p_0)$ is an analytic function of $p_0$.
\eq{QSR} is a general form of the QCD sum rule.
By properly parameterizing $\rho(p_0,\vecp)$, we obtain
QCD sum rules for physical quantities in $\rho(p_0,\vecp)$.

Let us first consider the spectral function, $\rho(p_0,\vecp)$.
The interpolating field couples to the states whose parity is opposite to that of the interpolating field,
as well as the states with the same parity of the interpolating field \cite{chung}.
Therefore, in the zero-width approxmation, \eq{3/2part} is expressed as
\begin{eqnarray}
\Pi(p)=\sum_n\left[|\lambda^n_-|^2{\psla+m^n_-\over p^2-{m^n_-}^2}
+|\lambda^n_+|^2{\psla-m^n_+\over p^2-{m^n_+}^2}\right],
\end{eqnarray}
where $m^n_{-,+}$ are the masses of negative and positive parity states,
$\lambda^n_{-,+}$ are the coupling strengths of the interpolating field with 
negative and positive parity states, respectively.
The spectral function in the rest frame, $\vecp={\bf 0}$, can be decomposed into two parts as follows,
\begin{eqnarray}\label{Spectral}
\rho(p_0)=
P_-\rho_{-}(p_0)+P_+\rho_{+}(p_0),
\end{eqnarray}
where $P_\mp=(\g_0\pm1)/2$ and $\rho_{\mp}(p_0)$ are given by
\bey
\rho_{\mp}(p_0)=\sum_n\left[|\lambda^n_\mp|^2\delta(p_0-m^n_\mp)+|\lambda^n_\pm|^2\delta(p_0+m^n_\pm)\right].
\label{spe-+}
%\rho_{-}(p_0)=\sum_n\left[|\lambda^n_-|^2\delta(p_0-m^n_-)+|\lambda^n_+|^2\delta(p_0+m^n_+)\right],
%\label{spe-}\\
%\rho_{+}(p_0)=\sum_n\left[|\lambda^n_-|^2\delta(p_0+m^n_-)+|\lambda^n_+|^2\delta(p_0-m^n_+)\right].
%\label{spe+}
\eey

Next, we construct the sum rule for negative parity states and that for positive parity.
We apply the projection operator $P_\mp$ to \eq{QSR} for $\vecp={\bf 0}$. Then we obtain
\begin{eqnarray}
\int_{-\infty}^{\infty}dp_0\rho_{\mp}^{\rm OPE}(p_0)W(p_0)=\int_{-\infty}^{\infty}dp_0\rho_{\mp}(p_0)W(p_0).
\label{projectQSR}
\end{eqnarray}
Note that in \eq{projectQSR} the contribution from the positive and negative parity states are not decoupled,
since, as can be seen from \eq{spe-+}, each of $\rho_{-}(p_0)$ and $\rho_{+}(p_0)$ contains the contribution from both of the parity states.
What we want to do is to separate the negative or positive parity contribution from Eqs.(\ref{projectQSR}).
(The following procedure is essentially equivalent to that in Ref.\cite{jido}.)

If $\rho_{\mp}^{\rm OPE}(p_0)$ are separable into $\rho_{\mp}^{\rm OPE}(p_0>0)$ and $\rho_{\mp}^{\rm OPE}(p_0<0)$,
we can separate \eq{projectQSR} into the contributions from
$p_0>0$ and $p_0<0$.
From the positive energy part, we obtain
\bey
\int_{0}^{\infty}dp_0\rho_{-}^{\rm OPE}(p_0)W(p_0)
=\int_{0}^{\infty}dp_0\rho_{-}(p_0)W(p_0),
\label{compprojectQSR-}\\
\int_{0}^{\infty}dp_0\rho_{+}^{\rm OPE}(p_0)W(p_0)
=\int_{0}^{\infty}dp_0\rho_{+}(p_0)W(p_0).
\label{compprojectQSR+}
\eey
Here we notice that only the negative (positive) 
parity states contribute to $\rho_{-}(p_0>0)$ ($\rho_{+}(p_0>0)$)
(see \eq{spe-+}).
\eq{compprojectQSR-} is therefore the sum rule for the negative parity states
and \eq{compprojectQSR+} is that for the positive parity states.

A comment is in order here.
In order to separate the negative and positive parity states in the sum rule as 
\eqs{compprojectQSR-}{compprojectQSR+},
it is necessary that $\rho_{\mp}^{\rm OPE}(p_0)$ are separable into $\rho_{\mp}^{\rm OPE}(p_0>0)$ and $\rho_{\mp}^{\rm OPE}(p_0<0)$
as mentioned above.
In general, $\rho_{\mp}^{\rm OPE}(p_0)$ are not separable \cite{kondo}.
However, as will be seen below (\eqst{ope}{A}{B}), $\rho_{\mp}^{\rm OPE}(p_0)$
for pentaquark is separable as long as we truncate the OPE at certain order,
since $\rho_{\mp}^{\rm OPE}(p_0)$ up to dimension 6 operator have the $p_0$ dependence as $p_0^n\left[\theta(p_0)-\theta(-p_0)\right]$.
We thus derive the sum rule for each parity state of the pentaquark as \eqs{compprojectQSR-}{compprojectQSR+}.

%\eq{compprojectQSR-} (\eq{compprojectQSR+}).
%\begin{eqnarray}
%&&\int_{0}^{\infty}dp_0\rho_{\mp}^{\rm OPE}(p_0)W(p_0)
%=\int_{0}^{\infty}dp_0\rho_{\mp}(p_0)W(p_0),\\
%&&\int_{-\infty}^{0}dp_0\rho_{\mp}^{\rm OPE}(p_0)W(p_0)
%=\int_{-\infty}^{0}dp_0\rho_{\mp}(p_0)W(p_0).
%\eey

We parameterize ${\rho_\mp}(p_0)$ with a pole plus continuum contribution,
\begin{eqnarray}\label{Phen}
\rho_\mp(p_0)=|\lambda_\mp|^2\delta(p_0-m_\mp)+|\lambda_\pm|^2\delta(p_0+m_\pm)\cr
\quad+[\theta(p_0-\omega_\mp)+\theta(-p_0-\omega_\pm)]\rho^{\rm OPE}(p_0).
\end{eqnarray}
Substituting \eq{Phen} into the right-hand sides of \eqs{compprojectQSR-}{compprojectQSR+},
we obtain the following sum rules,
\begin{eqnarray}\label{BSR}
\int_{0}^{\omega_\mp}dp_0\rho_\mp^{\rm OPE}(p_0)p_0^n\exp(-{p_0^2\over M^2})\cr
={m_\mp}^n|\lambda_\mp|^2\exp(-{{m_\mp}^2\over M^2})
\end{eqnarray}
Here we have chosen the weight function as $W(p_0)=p_0^n\exp(-p_0^2/M^2)$.
The parameter $M$ is called Borel mass.
%By doing this one can simultaneously improve the convergence of the OPE and 
%suppress the continuum contribution exponentially. 
From \eq{BSR} for $n=0$, we obtain the sum rule for the pole residues $|\lambda_\mp|^2$,
\bey
|\lambda_\mp|^2\exp(-{{m_\mp}^2\over M^2})=\int_{0}^{\omega_\mp}dp_0\rho_\mp^{\rm OPE}(p_0)\exp(-{p_0^2\over M^2}).
\label{residueSR}
\eey
The ratio of \eq{BSR} for $n=0$ and $n=2$ gives the sum rules for the masses,
\bey
\left(m_{\mp}\right)^2=\frac{\int_{0}^{\omega_\mp}dp_0\rho_\mp^{\rm OPE}(p_0)p_0^2\exp(-{p_0^2\over M^2})}{\int_{0}^{\omega_\mp}dp_0\rho_\mp^{\rm OPE}(p_0)\exp(-{p_0^2\over M^2})}.
\label{massSR}
\eey

Let us now turn to the OPE.
We have taken into account the terms up to dimension 6 operator.
We show the result of the OPE,
\bey
\rho^{\rm OPE}(p_0)&=&\g_0 A(p_0)+B(p_0),
\label{ope}
\eey
where $A(p_0)$ and $B(p_0)$
are given by
\bey
A(p_0)&=&
\left[\frac{1}{5^2\cdot 3^2 \cdot 2^{18}\pi^8}(p_0)^{11}
\right.\cr
&&
+\frac{-1}{5\cdot 3^4\cdot 2^{17}\pi^6}\GG
(p_0)^7
\cr
&&
+\frac{1}{3^3\cdot 2^{12}\pi^6}m_{s}\sc
(p_0)^{7}\cr
&&
+\frac{-1}{5\cdot3^2\cdot 2^{10}\pi^6}
m_{s}\sGs
%g\langle{\bar s}\sigma\cdot{\cal G}s\rangle 
(p_0)^{5}\cr
&&\left.
+\frac{1}{5\cdot 2^{8} \pi^4}
\qcond^2
%\bra\bar{q} q\ket^2
(p_0)^{5}
%+\frac{5}{3^3\cdot 2^5 \pi^2}m_s \sc \bra\bar{q} q\ket^2 p_0
\right]\cr
&&\times\left[\theta(p_0)-\theta(-p_0)\right],
\label{A}\\
B(p_0)&=&
\left[
\frac{1}{7\cdot 5^2\cdot 3\cdot 2^{14}\pi^8}m_s (p_0)^{10}
\right.\cr
&&
+\frac{-1}{3^3\cdot 2^{13}\pi^6}\sc(p_0)^8
\cr
&&
\left.
+\frac{1}{5\cdot 3^2\cdot 2^{10}\pi^6}
\sGs
%g\langle{\bar s}\sigma\cdot{\cal G}s\rangle
(p_0)^6
\right]\cr
&&\times\left[\theta(p_0)-\theta(-p_0)\right].
\label{B}
\eey
In \eqs{A}{B}, $q=u, d$, $m_s$ is the strange quark mass, and $\bra 0|{\cal O}|0\ket$ denotes
the vacuum expactation value of the operator ${\cal O}$.

Here, before deriving the QCD sum rules for $\Theta^{++}$, we comment on 
the contribution of the continuum states. In Ref.\cite{kmn}, it was pointed out that pentaquark correlation functions receive contribution of two-hadron-reducible (2HR) diagrams, 
which represent baryons and mesons propagating independently without interacting with each other.
The 2HR diagrams are related only with the background (continuum states).
We can make the background contribution in the sum rules as small as possible by subtracting the 2HR diagrams.
It is better to subtract them especially in the sum rules for $J^P=1/2^{\pm}$ pentaquark, where the $NK$ continuum contribution should be significant.
However, in the sum rules for $J^P=3/2^{-}$ pentaquark, the $NK$ continuum contribution itself is expected to be small, since the $N$ and $K$ are relatively $D$-wave in this channel.
Hence, in this paper, we consider the correlation function without subtracting the 2HR parts.

We substitute $\rho_\mp^{\rm OPE}(p_0)=A(p_0)\pm B(p_0)$ with \eqs{A}{B} into the right hand sides of
\eqs{residueSR}{massSR}.
Then we obtain the QCD sum rules for $\Theta^{++}$.

\begin{figure}
%\rotatebox{-90}
{\includegraphics[width=8cm,keepaspectratio]{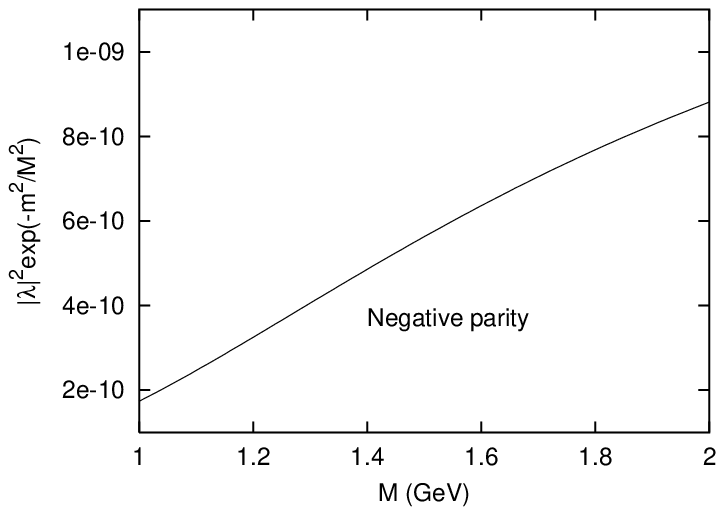}}
\caption{$|\lambda_-|^2\exp(-{{m_-}^2\over M^2})$ as functions of Borel mass, $M$ with the continuum threshold parameter $\omega_-=1.8\,{\rm GeV}$.}
\label{res-}
\end{figure}

We plotted in Fig.\ref{res-} the right-hand side of \eq{residueSR} for the negative parity state as a function of the Borel mass, $M$.
Here and hereafter we use the standard values of the QCD parameters, 
$\qcond=(-0.23\;{\rm GeV})^3$,
$m_s=0.12\;{\rm GeV}$, $\sc=0.8\qcond$,
$\sGs=(0.8\;{\rm GeV}^2)\sc$, $\GG=(0.33\;{\rm GeV})^4$.
%The solid and dashed lines are for the negative and positive parity states, respectively.
As can be seen, the right-hand side of \eq{residueSR} is positive.
This implies that the present QCD sum rule does not exclude
the possiblity of the exisitence of the $\Theta^{++}$ with negative parity
since the left-hand side of \eq{residueSR} must be positive.

\begin{figure}
%\rotatebox{-90}
{\includegraphics[width=8cm,keepaspectratio]{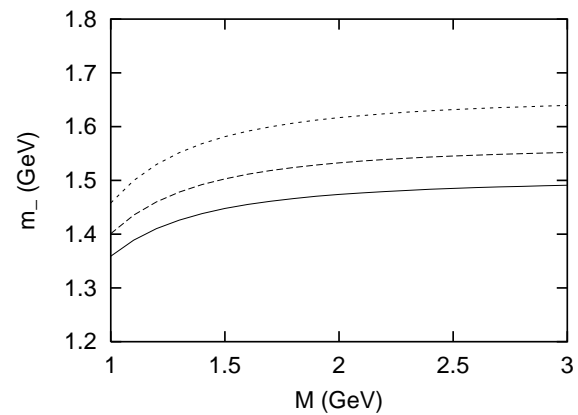}}
\caption{Mass of $J^P=3/2^-$, $I=1$ pentaquark as a function of Borel mass, $M$, with the continuum threshold parameters $\omega_{-}=1.73\,{\rm GeV}$ (solid line), $1.8\,{\rm GeV}$ (dashed), $1.9\,{\rm GeV}$ (dotted).}
\label{mass-}
\end{figure}

In Fig.\ref{mass-}, we plotted the mass of $\Theta^{++}$ with negative parity against the Borel mass which is obtained from \eq{massSR}.
We see that the dependence on the Borel mass is weak.
This implies that the sum rule works well.
However, the result depends on the choice of the continuum thershold $\omega_-$.
The continuum mainly comes from the $S$-wave $\Delta K$ scattering states,
whose threshold is $1.73\,{\rm GeV}$,
since $NK$ scattering state must be $D$-wave in this channel and 
it starts up very gradually.
Thus we choose $\omega_{-}=1.73 $, $1.8$, $1.9\,{\rm GeV}$.
From the stabilized region of the curve,
we predict the mass to be $ 1.5\sim 1.6$ GeV,
which is close to the observed $\Theta^+$ mass.
The mass is much below the $\Delta K$ threshold.
This implies that the $\Theta^{++}$ with $J^P=3/2^-$ can be extremely narrow
since it is allowed to decay only to $D$-wave $KN$ state
and the width is strongly suppressed due to the large centrifugal barrier.
%If the pentaquark lies above the $\Delta K$ threshold,
%the $\Delta K$ decay channel opens and the width is large.
%It should be noted that if the mass is below but close to the $\Delta K$ threshold it cannot be narrow since the $\Delta$ has a large width ($\sim 120\,{\rm MeV}$).

%The predicted mass of $J^P=3/2^-$, $I=1$ pentaquark is close to that of $\Theta^+$.
It is remarkable that such a high spin and isovector state can be a low
lying state, which is not the case for ordinary baryons.
The possibility of $J^P=3/2^-$, $I=1$ pentaquark being low lying state
%The mechanism for that 
%The mechanism for such a high spin and unnatural parity state being low lying state 
has been suggested 
%has been elucidated 
by the previous calculation from a quark model \cite{amdpenta}.
In Ref.\cite{amdpenta}, a simple quark model
in which constituent quarks interact via one-gluon exchange force at short distances and confining (or string) potential at long distances was considered.
A $qqqq\bar{q}$ system has a connected string configuration corresponding to a confined state, in addition to an ordinary meson-baryon like configuration.
%Assuming the latter is decoupled,
A variational method called antisymmetrized
molecular dynamics (AMD) \cite{amd1,amd2} was applied to the confined $uudd\bar{s}$ system
and all the possible spin parity states were calculated. 
The narrow and low lying states they have found are $J^P=1/2^+$ or $3/2^+$ with $I=0$
and $J^P=3/2^-$ with $I=1$ states.
The former has just the same structure as that conjectured by Jaffe and Wilczek \cite{jaffe}. 
We represent it as $[ud]_{S=0,I=0}[ud]_{S=0,I=0}[\bar{s}]$,
where $[ud]_{S,I}$ denotes a color $\bf\bar 3$ $ud$-diquark with spin $S$ and isospin $I$. 
Both of the two diquarks gain color magnetic interaction since they have $S=0$. However, this state loses the kinetic and string energy,   
since the two diquarks, which are to be antisymmetric in color, are identical and must be relatively $P$-wave.
In Ref.\cite{amdpenta}, another energetically favorable state
has been predicted,
which consists of an $S=0$ diquark and an $S=1$ diquark:
$[ud]_{S=0,I=0}[ud]_{S=1,I=1}[\bar{s}]$.
%The interpolating field used in the present QCD sum rule 
%has the same diquark structure. 
This state is totally $J^P=3/2^-$ and $I=1$.
It loses color magnetic interaction since one of the diquarks has $S=1$.
However, it gains kinetic and string energy, since the two diquarks are no longer identical and they can be relatively $S$-wave.
Owing to the balance between the energy gain and loss, $J^P=3/2^-$, $I=1$ state degenerate with $J^P=1/2^+$ or $3/2^+$, $I=0$ state.
Within the quark model employed in Ref.\cite{amdpenta}, however, 
one cannot predict the absolute masses but only the level structure of the pentaquarks, because this quark model relies on the 
zero-point energy of the confining potential.
In Ref.\cite{amdpenta}, it was adjusted to reproduce the observed mass of $\Theta^+$.
Whereas, the QCD sum rule is able to estimate the absolute mass.
We confirmed from the QCD sum rule that the $J^P=3/2^-$, $I=1$ state actually
can be a low lying state, 
using the interpolating field, \eq{current}, which has the same structure as that suggested by the quark model.
%The quark model calculation confirmed it actually does.
%$J^P=3/2^-$, $I=1$ pentaquark is a state newly predicted by us,
%which will be a candidate of $\Theta^{++}$.
%which consists of a spin-0 diquark, a spin-1 diquark and an anti-strange %quark.
%The two diquarks are relatively s-wave.
%A remarkable feature is that they can be almost degenerate due to the %balance of the loss of the kinetic and string energies with the gain of the %color magnetic interaction. 

The pentaquark with $J^P=3/2^-$ and $I=1$ has also been found from the chiral unitary approach, as a resonance state in the $\Delta K$ channel \cite{oset}.
This state is generated due to an attractive interaction in that channel existing in the lowest order chiral Lagrangian.
The attractive interaction leads to a pole of the complex energy plane and manifests itself in a large strength of the $\Delta K$ scattering amplitude with $L=0$ and $I=1$.
We note that the interpolating field, \eq{current}, can also couple 
with such a $\Delta K$ resonance states 
because it contains the $\Delta K$ component as is shown by Fierz transformation.
%We note that the corresponding pentaquark state predicted from the present QCD sum rule also contains the $\Delta K$ component,
%because the interpolating field, \eq{current}, can be expressed as a sum of products
%of baryon and meson interpolating fields including a product of $\Delta$ and $K$, after Fierz transformation. 

\begin{figure}[t!]
%\rotatebox{-90}
{\includegraphics[width=8cm,keepaspectratio]{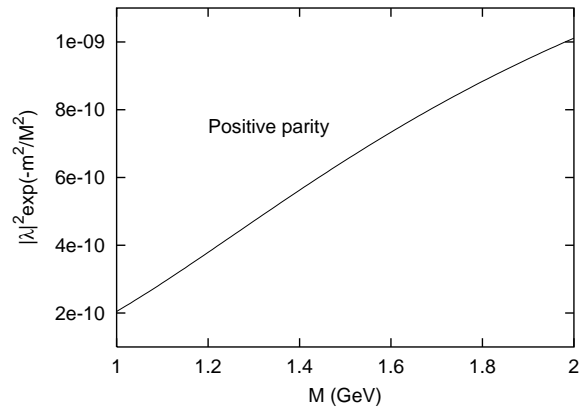}}
\caption{$|\lambda_+|^2\exp(-{{m_+}^2\over M^2})$ as functions of Borel mass, $M$ with the continuum threshold parameter $\omega_+=1.8\,{\rm GeV}$.}
\label{res+}
\end{figure}

\begin{figure}[h!]
%\rotatebox{-90}
{\includegraphics[width=8cm,keepaspectratio]{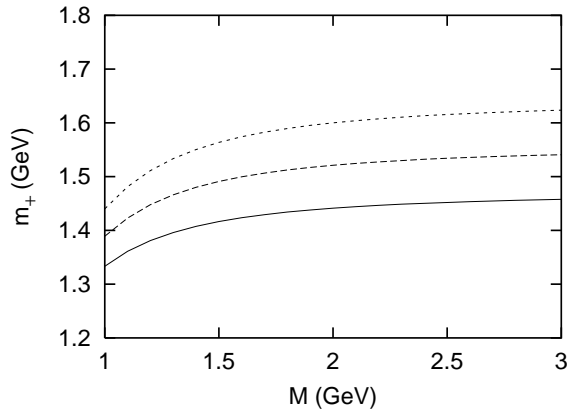}}
\caption{Mass of $J^P=3/2^+$, $I=1$ pentaquark as a function of Borel mass, $M$ with the continuum threshold parameters $\omega_{+}=1.7\,{\rm GeV}$ (solid line), $1.8\,{\rm GeV}$ (dashed), $1.9\,{\rm GeV}$ (dotted).}
\label{mass+}
\end{figure}

Let us turn to the sum rule for the positive parity state.
We plotted in Fig.\ref{res+} the right-hand side of \eq{residueSR} for the positive parity state as a function of the Borel mass, $M$.
The right-hand side of \eq{residueSR} is positive, which implies that
the exisitence of the positive parity state is not excluded.
The mass against the Borel mass is shown
in Fig.\ref{mass+}.
The continuum in this channel mainly comes from the $P$-wave $NK$ scattering states.
We choose $\omega_{+}=1.7 $, $1.8$, $1.9\,{\rm GeV}$.
Although the curve depends on the choice of the continuum threshold parameter, we can say that the positive parity state nearly degenerate with the negative parity state.
However, the positive parity state is expected to be broader than the negative parity state,
since the former can decay into $P$-wave $NK$ states while 
the latter only to $D$-wave $NK$ states.
The present result is consistent with a recent calculation by Skyrme model \cite{borisyuk}.
The authors in Ref.\cite{borisyuk} predicted that there exists a new isotriplet
of $\Theta$-baryons with $J^P=3/2^+$ and $I=1$.
Its mass is $1595\,{\rm MeV}$ and the width is large: $\Gamma\sim 80\,{\rm MeV}$.


In summary, we have studied $J=3/2$, $I=1$ pentaquark, $\Theta^{++}$, using the method of QCD sum rule.
We used the interpolating field constructed from a color anti-triplet scalar isoscalar diquark, a color anti-triplet axial-vector isovector diquark
and an anti-strange quark.
We have derived the QCD sum rules for the negative and positive parity states.
QCD sum rule predicts a narrow $\Theta^{++}$ ($J^P=3/2^-$).
Its mass is predicted to be $1.5\sim 1.6$ GeV,
which is much below the $\Delta K$ threshold.
Since only the $D$-wave decay to $NK$ channel is allowed,
it should be an extremely narrow state.
%, whose width is $\Gamma<1\,{\rm MeV}$.
%Thus, QCD sum rule supports the results of our previous calculation using
%a quark model \cite{amdpenta}. 
QCD sum rule also shows the possibility of the existence of 
the $J^P=3/2^+$ state.
It nearly degenerates with the negative parity state.
It may be broader than the negative parity state,
since it is allowed to decay into $P$-wave $NK$ state.
It is worth mentioning that this is the first QCD sum rule analysis of 
high spin states of the pentaquark.
Most of the works using QCD sum rules and Lattice QCD 
concentrate on $J=1/2$ and $I=0$ pentaquark states.
It would be interesting to see if lattice calculation could confirm these
findings.

The existence of the $\Theta^{++}$ has not been experimentally confirmed yet.
In search of a particle, one should pay attention to its properties, because
the production rates depend on the spin, parity and width of the particle.
We would like to stress that the $\Theta^{++}$ with $3/2^-$ may be extremely narrow. 
Therefore, it would be helpful to carefully choose the entrance channels in the $\Theta^{++}$ search. 

Further observations of pentaquark states with various quanta are requested to
give insight into the structure of the multiquark systems, and would lead to a deeper underestanding of exotic hadrons.

%%%%%%%%%%%%%%%%%%%%%%%%%%%%%%%%%%%%%%%%%%%%%%%%%%%%%%%%%%%%%%%%%

%\baselineskip 24pt
%\begin{center}
%{\bf References}
%\end{center}
%\def\labelenumi{[\theenumi]}
%%%%%%%%%%%%%%%%%%%%%%%%%%%%%%%%%%%%%%%%%%%%%%%%%%%%%%%%%%%%%%%%%
\def\Ref#1{[\ref{#1}]}
\def\Refs#1#2{[\ref{#1},\ref{#2}]}
\def\npb#1#2#3{{Nucl. Phys.\,}{\bf B{#1}},\,#2\,(#3)}
\def\npa#1#2#3{{Nucl. Phys.\,}{\bf A{#1}},\,#2\,(#3)}
\def\np#1#2#3{{Nucl. Phys.\,}{\bf{#1}},\,#2\,(#3)}
\def\plb#1#2#3{{Phys. Lett.\,}{\bf B{#1}},\,#2\,(#3)}
\def\prl#1#2#3{{Phys. Rev. Lett.\,}{\bf{#1}},\,#2\,(#3)}
\def\prd#1#2#3{{Phys. Rev.\,}{\bf D{#1}},\,#2\,(#3)}
\def\prc#1#2#3{{Phys. Rev.\,}{\bf C{#1}},\,#2\,(#3)}
\def\prb#1#2#3{{Phys. Rev.\,}{\bf B{#1}},\,#2\,(#3)}
\def\pr#1#2#3{{Phys. Rev.\,}{\bf{#1}},\,#2\,(#3)}
\def\ap#1#2#3{{Ann. Phys.\,}{\bf{#1}},\,#2\,(#3)}
\def\prep#1#2#3{{Phys. Reports\,}{\bf{#1}},\,#2\,(#3)}
\def\rmp#1#2#3{{Rev. Mod. Phys.\,}{\bf{#1}},\,#2\,(#3)}
\def\cmp#1#2#3{{Comm. Math. Phys.\,}{\bf{#1}},\,#2\,(#3)}
\def\ptp#1#2#3{{Prog. Theor. Phys.\,}{\bf{#1}},\,#2\,(#3)}
\def\ib#1#2#3{{\it ibid.\,}{\bf{#1}},\,#2\,(#3)}
\def\zsc#1#2#3{{Z. Phys. \,}{\bf C{#1}},\,#2\,(#3)}
\def\zsa#1#2#3{{Z. Phys. \,}{\bf A{#1}},\,#2\,(#3)}
\def\intj#1#2#3{{Int. J. Mod. Phys.\,}{\bf A{#1}},\,#2\,(#3)}
\def\sjnp#1#2#3{{Sov. J. Nucl. Phys.\,}{\bf #1},\,#2\,(#3)}
\def\pan#1#2#3{{Phys. Atom. Nucl.\,}{\bf #1},\,#2\,(#3)}
\def\app#1#2#3{{Acta. Phys. Pol.\,}{\bf #1},\,#2\,(#3)}
\def\jmp#1#2#3{{J. Math. Phys.\,}{\bf {#1}},\,#2\,(#3)}
\def\cp#1#2#3{{Coll. Phen.\,}{\bf {#1}},\,#2\,(#3)}
\def\epjc#1#2#3{{Eur. Phys. J.\,}{\bf C{#1}},\,#2\,(#3)}
\def\mpla#1#2#3{{Mod. Phys. Lett.\,}{\bf A{#1}},\,#2\,(#3)}
\def\etal{{\it et al.}}
%%%%%%%%%%%%%%%%%%%%%%%%%%%%%%%%%%%%%%%%%%%%%%%%%%%%%%%%%%%%%%

%%%%%%%%%%%%%%%%%%%%%%%%%%%%%%%%%%%%%%%%%%%%%%%%%%%%%%%%%%%%%%%%%%%%
%\divide\baselineskip by 4
%\multiply\baselineskip by 3

%%%%%%%%%%%%%%%%%%%%%%%%%%%%%%%%%%%%%%%%%%%%%%%%%

\end{document}

%% file: def.tex
\def\ben{\begin{equation}}
\def\een{\end{equation}}
\def\bey{\begin{eqnarray}}
\def\eey{\end{eqnarray}}
\def\ba{\begin{array}}
\def\ea{\end{array}}
\def\benmrt{\begin{enumerate}}
\def\eenmrt{\end{enumerate}}
\def\psla{p{\raise1pt\hbox{$\!\!\!/$}}}
\def\dsla{\partial{\raise1pt\hbox{$\!\!\!/$}}}
\def\Dsla{D{\raise1pt\hbox{$\!\!\!/$}}}
\def\xsla{x{\raise1pt\hbox{$\!\!\!/$}}}
\def\jmu5{j_{\mu 5}^{(i)}(0)}
\def\jnu5{j_{\nu 5}^{(i)}(0)}
\def\qq0v{\langle0\!\mid\!{\bar q}q\!\mid\! 0\rangle}

\def\qc0f{\langle0\!\mid\!{\bar q}q\!\mid\!0\rangle_{F}}

\def\qsq0f{\langle0\!\mid\!{\bar q}\sigma_{\mu\nu}q\!\mid\!0\rangle_{F}}

\def\qgdq0f{\langle0\!\mid\!{\bar q}{\cal 
S}\gamma_{\mu}D_{\nu}q\!\mid\!0\rangle_{F}}

\def\qddq0f{\langle0\!\mid\!{\bar q}{\cal
S}D_{\mu}D_{\nu}q\!\mid\!0\rangle_{F}}
\def\3mmtm{|{\bf q}|^2}

\def\eq#1{Eq.(\ref{#1})}
\def\eqs#1#2{Eqs.(\ref{#1}) and (\ref{#2})}
\def\eqst#1#2#3{Eqs.(\ref{#1}), (\ref{#2}) and (\ref{#3})}

\def\Ref#1{[\ref{#1}]}
\def\Refs#1#2{[\ref{#1},\ref{#2}]}

\def\p0{p_0}

\def\gam3{\mbox{\boldmath{$\gamma$}}}
\def\e0{E_{0}(s_{0},s)}
\def\e1{E_{1}(s_{0},s)}
\def\e2{E_{2}(s_{0},s)}

\def\aplt{\kern0.3333em \raise 0.2ex \hbox{$<$}%
\kern-0.8em \lower0.8ex \hbox{$\sim$}%
\kern0.3333em}
\def\aplg{\kern0.3333em \raise 0.2ex \hbox{$>$}%
\kern-0.8em \lower0.8ex \hbox{$\sim$}%
\kern0.3333em}